\documentclass[%
 reprint,
showpacs,preprintnumbers,
 amsmath,amssymb,
 aps,
prb,
]{revtex4-1}
\usepackage[english]{babel}
\usepackage{graphicx}% Include figure files
\usepackage{dcolumn}% Align table columns on decimal point
\usepackage{bm}% bold math
\usepackage[version=3]{mhchem}

\begin{document}

\title{Local removal of silicon layers on Si(100)-2$\times$1 with chlorine-resist STM lithography}

\author{T. V. Pavlova$^{1,2}$}
\email{pavlova@kapella.gpi.ru}
\author{V. M. Shevlyuga$^{1}$}
\author{B.~V.~Andryushechkin$^{1,2}$}
\author{G.~M.~Zhidomirov$^{1,2}$}
\author{K. N. Eltsov$^{1,2}$}
\affiliation{$^{1}$Prokhorov General Physics Institute of the Russian Academy of Sciences, Vavilov str. 38, 119991 Moscow, Russia}
\affiliation{$^{2}$National Research University Higher School of Economics, Myasnitskaya str. 20, 101000 Moscow, Russia}

\begin{abstract}

We report the realization of STM-based lithography with silicon layers removal on the chlorinated Si(100)-2$\times$1 surface at 77\,K. In contrast to other STM lithography studies, we were able to remove locally both chlorine and silicon atoms. Most of the etched pits have a lateral size of 10--20\,{\AA} and a depth of 1--5\,{\AA}. In the pits in which the STM image with atomic resolution is obtained, the bottom is mainly covered with chlorine. Some pits contain chlorine vacancies. Mechanisms of STM-induced removal of silicon and chlorine atoms on Si(100)-2$\times$1-Cl are discussed and compared with the well-studied case of STM-induced hydrogen desorption on Si(100)-2$\times$1-H. The results open up new possibilities of the three-dimensional local etching with STM lithography.

\end{abstract}

\maketitle

\section{Introduction}

The ability of scanning tunneling microscope (STM) to manipulate atoms on surfaces and to create artificial nanostructures has stimulated a great number of research works since the 1990s. STM nanolithography on silicon surfaces attracted more attention due to their technological importance.

STM lithography on hydrogen-passivated Si(100) is one of well-developed technologies used for the fabrication of nanodevices. This approach assumes the use of hydrogen monolayer as a resist, which can be easily removed with an STM tip \cite{1995Shen, 2009Walsh, 2017Moller, 2018Achal}. Recently, the hydrogen depassivation lithography has evolved into a well-established technique applied with success to the creation of a single atom transistor \cite{2012Fuechsle} and elements of the quantum computer  \cite{2018Hileeaaq, 2018Broome}.

Apart from hydrogen-passivated silicon surfaces, there are alternative resists that can be used with STM lithography for the fabrication of nanodevices (see, for example, Ref. \cite{2009Walsh}). In particular, a resist can be formed by chemisorbed halogen atoms  \cite{1993Boland, 2003Nakamura, 2007Nakamura, 1994Baba64, 1994Baba12, 2018Butera}. The manipulation of a halogen atom on a silicon surface has a long history. In particular, Boland \cite{1993Boland} reported the switching of atomic positions of chlorine atoms on Si(100) under the influence of an STM tip. Nakamura et al. \cite{2003Nakamura, 2007Nakamura} reported observation of chlorine atom diffusion and desorption induced by carrier injection from an STM tip. Baba and Matsui \cite{1994Baba64, 1994Baba12} reported the selective atomic desorption of chlorine atoms from Si(111)-7$\times$7 surface. Dwyer et al. \cite{2018Butera} demonstrated the STM lithography on the Si(100)-2$\times$1-Cl surface with Cl atoms removal.

Although STM lithography on chlorine-passivated silicon surfaces is of great interest as a direct alternative to hydrogen depassivation lithography, there is another potential application determined by the capability of chlorine to etch a silicon crystal. If chlorine-induced STM lithography can remove not only adsorbate but also silicon atoms, we could create the local sites to insert necessary foreign atoms directly into the silicon lattice. This new task comes from recent theoretical study \cite{2018Pavlova}, in which the Si(100)-2$\times$1-Cl system was considered for the optimization of the precise placement of a single phosphorus atom into the silicon lattice. In particular, the authors demonstrated that a single silicon vacancy on the Si(100)-2$\times$1-Cl surface is a perfect site for P atom incorporation as a result of PH$_{3}$ adsorption. What remained was to learn how to extract a single Si atom from the substrate together with one Cl atom keeping the chlorine resist unchanged.

There are several reports concerned the extraction of silicon atoms from a halogenated silicon surface. In particular, Baba and Matsui \cite{1994Baba65} observed movement and desorption of a silicon atom from a chlorinated Si(111)-7$\times$7 surface. Atomic-layer etching of Br-saturated Si(111) surfaces was achieved by using a scanning tunneling microscope at room temperature in the scanning mode \cite{1999Mochiji}. In another report on electron-stimulated desorption from Br-chemisorbed Si(111)-7$\times$7 surface \cite{2000Mochiji}, it was found out that irradiation of a surface by field emission electrons from an STM tip leads to the desorption of Si adatoms as well as Br atoms.

In this work, we have realized STM lithography with Si atoms removal from a chlorinated Si(100)-2$\times$1 surface. We demonstrate that at proper conditions both chlorine and silicon atoms desorb from the surface producing etched pits on the Si substrate. To confirm that one or more silicon layers and some chlorine atoms are missing in the etch pits, we perform a detailed analysis of the pits. Our work also contains theoretical analysis of possible mechanisms behind the extraction of chlorine and silicon atoms.

\section{Methods}

\subsection{Experimental methods}

All experiments were carried out in a multi-chamber ultrahigh vacuum (UHV) setup with Low-Temperature Scanning Tunneling Microscope GPI CRYO operating at 77\,K. During the experiments, the base pressure was better than  5$\cdot$10$^{-11}$\,Torr. The Si(100) samples were cut from a p-type boron-doped Si wafer (1\,$\Omega \cdot$\,cm). The sample with the surrounding armature was outgassed for three days in UHV, keeping the temperature of the sample at 1070\,K.

The low-defect Si(100)-2$\times$1-Cl surface for the experiment was first prepared by flash heating up to 1470\,K for 5\,s to remove the oxide, then by adsorbing molecular chlorine injected through a fine leak piezo-valve at partial pressure of 10$^{-8}$\,Torr  during 100--200\,s. The introduction of molecular chlorine was done approximately at sample temperature of 370--420\,K just after turning off the flash heating.

Polycrystalline tungsten tips were used both for STM imaging and for lithography. The tips were manufactured by electrochemical etching 0.25\,mm diameter polycrystalline tungsten wire in 2\,M NaOH solution. For cleaning and sharpening, the tips were bombarded in UHV with Ar$^+$ ions. To improve the tip stability for the lithography experiments, the tips were subsequently annealed at 2070--2270\,K.

\subsection{Computational methods}

Spin-polarized density functional theory (DFT) calculations were performed with the generalized-gradient approximation (GGA) according to Perdew, Burke, and Ernzerhof (PBE) \cite{1996Perdew} implemented in VASP code  \cite{1993Kresse, 1996Kresse}.  Semi-empirical Grimme’s DFT-D2 dispersion correction \cite{2006Grimme} was applied for all calculations. The Si(100)-2$\times$1 surface was simulated by recurring 4$\times$4 cells, each consisting of eight atomic layers of silicon. A vacuum space of approximately 15\,{\AA} was introduced between the slabs. The bottom three layers were fixed in bulk positions, while the other silicon layers were allowed to relax. The lower side of the slab was covered by hydrogen atoms to saturate the dangling bonds of silicon. Chlorine atoms were placed on the upper side of the slab as a 2$\times$1-Cl lattice. Reciprocal cell integrations were performed using the 4$\times$4$\times$1 k-points grid.  STM images were simulated in the framework of Tersoff-Hamann approximation \cite{1985Tersoff}.

The phonon frequencies were calculated using the density functional perturbation theory (DFPT), as implemented in VASP. For that, Si(100)-2$\times$1-Cl and Si(100)-2$\times$1-H surfaces were simulated by  2$\times$2 cells with eight atomic layers of silicon (the bottom three layers were fixed and the lower side of the slab was covered by hydrogen). Reciprocal cell integrations were performed with the 8$\times$8$\times$1 k-points grid. For bonding analysis, the electronic density of states (DOS) and the crystal orbital Hamilton population (COHP) \cite{1993Dronskowski} was evaluated using LOBSTER \cite{2011Deringer, 2013Maintz, 2016Maintz}. For DOS and COHP calculations, 2$\times$1 cell was used with the 10$\times$20$\times$1 k-points grid.

\section{Results and discussion}

\subsection{Pit creation and characterization}

The pits on the Si(100)-2$\times$1-Cl surface were created according to the following procedure. First, the tip was positioned over the desired location on the surface under the tunnel gap $U_s = +2$\,V, $I_{t}=1$\,nA ($U_s$ is the sample voltage), which reduces the tip-sample distance compared to that under our typical scanning conditions ($U_s = +4$\,V, $I_{t}=1$\,nA). After that the feedback was turned off and the sample was supplied with initial voltage ($U_s^i \approx +3$\,V), increasing linearly to critical voltage ($U^c_s \approx +4$\,V), and turned back linearly to $U_s^i$. A voltage changes stepwise with a step of 0.1\,V and a speed of 1\,V/s. Therefore, the total exposure time was about 2\,s. We determine the critical voltage as the voltage, at which minimal defect is observed on the surface (we define the minimum defect as a defect of about 2$\times$2 silicon atoms, since single vacancies were very rare in our experimental conditions). While a small decrease in $U^c_s$ by a value of 0.1--0.2\,V led to the absence of the pits, increasing it by the same value led to a significant increase in the size of the pits (up to 10\,nm). The critical voltage depends on the tip-sample distance, which can be adjusted by the initial current when positioning the tip before the voltage pulse. The critical voltage decreases to $\approx 3$\,V when the tip approaches the surface by $\approx 1.5$\,{\AA}  (at the initial current $I^i_{t}=5$\,nA instead of $I^i_{t}=1$\,nA) and $U^c_s$ increases to $\approx 5$\,V when the tip moves away from the surface by $\approx 1.5$\,{\AA} (at $I^i_{t}=0.2$\,nA).

\begin{figure}
\begin{center}
    \includegraphics[width=0.92\linewidth]{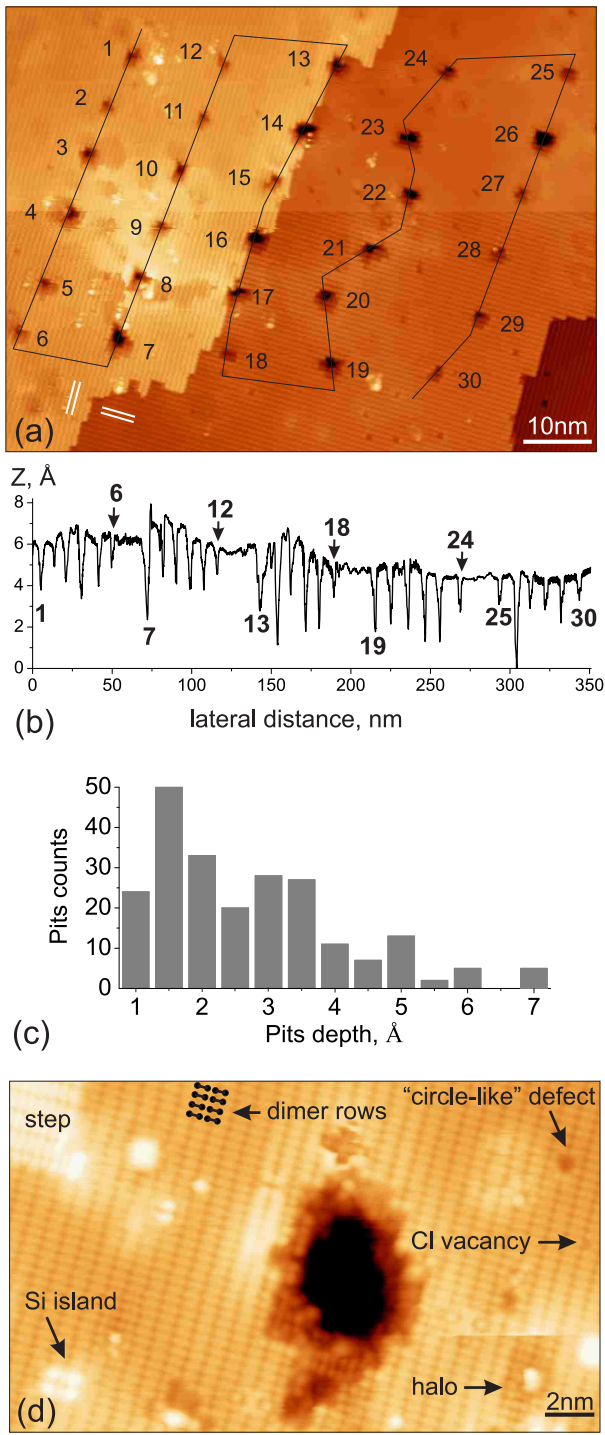}
    \caption{(a) Filled state STM image ($U_s =-4.0$\,V, $I_t = 1.0$\,nA) of the  5$\times$6 array of nanopits created on Si(100)-2$\times$1-Cl. Pits were created on two atomic terraces, dimer rows directions on terraces are highlighted by pairs of white parallel lines. Pits were created by a voltage pulse ramped from $+3$ to $+4$\,V and back with total duration of 2\,s (details see in the text). (b) Line-scan across the nanopits array. (c) Histogram of the depths of the pits (based on the analysis of the STM images of 225 pits). (d)  Empty state STM image ($U_s =+2.5$\,V, $I_t = 1.9$\,nA) of the Si(100)-2$\times$1-Cl surface with defects after the pit creation. Silicon dimers are marked by black dumb-bells.}
    \label{fig1}
\end{center}
\end{figure}

Figure~\ref{fig1}a shows an array 5$\times$6 of the etch pits fabricated on the Si(100)-2$\times$1-Cl  surface. The pits have a lateral dimension of 1--2\,nm, which approximately corresponds to two dimer rows each consisting of four dimers. Keeping the same parameters (tunnel current, voltage, exposure time), we were able to create up to ten pits with the same probability. After that, the efficiency of pit creation became lower, which required slightly increasing the threshold voltage or exposure time to recover the patterning. This effect can be explained by the change of the tip apex state as a result of the tip-surface interaction \cite{2017Moller, 2003Soukiassian1}, which was also manifested as lower contrast and spatial resolution of STM images recorded after the creation of a pit. (One possible reason is the deposition of silicon from the pit onto the tip). Note also that for different W-tips (prepared in the same way), as well as for different states of the tip, the optimal parameters necessary for reproducible creation of pits differed by $\approx 0.5$\,V.

Figure~\ref{fig1}b shows an STM line-scan across the pits in Fig.~\ref{fig1}a. The distribution of the pits with respect to their depth is represented in Fig.~\ref{fig1}c. Note that it is difficult to calculate the correct depth of the pit, because the spatial resolution of the tip degrades after creating the pit, and this affects the measurement of the depth of the pit. A thick tip underestimates the depth of the pit. We roughly estimate the measurement error as $\approx 0.5$\,{\AA}. The depth of the steps on the Si(100)-2$\times$1-Cl surface is 1.4\,{\AA}, while the depth of the chlorine vacancy is about 0.8--1.4\,{\AA}. From the analysis of 225 pits, we have found that only 24 pits have heights of $\approx 1$\,{\AA}, it means that approximately 11\% of the pits may not be deep enough and contain only chlorine vacancies. The remaining pits are deeper, their depth is about 1.5--5.0\,{\AA} and only 5\% of the pits are deeper than 5\,{\AA}. Therefore, we believe that the removal of silicon occurs at least in 89\% of the pits.

Figure~\ref{fig1}d shows defects on the surface after the pit creation. The Si(100)-2$\times$1-Cl system is formed by Si dimers and chlorine atoms bound to each Si atom of dimer (see, for example, Ref. \cite{2004Lee}). Initial concentration of atomic defects on the Si(100)-2$\times$1-Cl surface did not exceed $0.2\%$. Most defects are single chlorine vacancies and "circle-like" defects (Fig.~\ref{fig1}d) (to our knowledge, the structure of the "circle-like" defect has not been discussed anywhere). After creating the pit, the surface area around the pit became more defective. Defects studied on the Si(100)-2$\times$1-Cl surface \cite{2006Trenhaile} include only single and double Cl vacancies, bare dimers, and split dimers, therefore we cannot identify all the defects in Fig.~\ref{fig1}d. Possible defects can contain Si and Cl atoms removed from the pits. A halo near a charged defect looks similar to those on the hydrogenated Si(100)-2$\times$1 surface around negatively charged H vacancy \cite{2015Labidi}, thus this defect can be a negatively charged Cl vacancy.  Si island of three dimers is visible in the lower left corner of STM image in Fig.~\ref{fig1}d (the distances between neighboring atoms of the island are equal to those on the Si(100)-2$\times$1-Cl surface, and the height of the island is equal to the height of the step). Note that much more silicon islands should be observed if silicon removed from the pits is adsorbed to the surface. However, we did not observe them, therefore, we believe that silicon removed from the pits is deposited on the tip.

\begin{figure*}
\begin{center}
    \includegraphics[width=\linewidth]{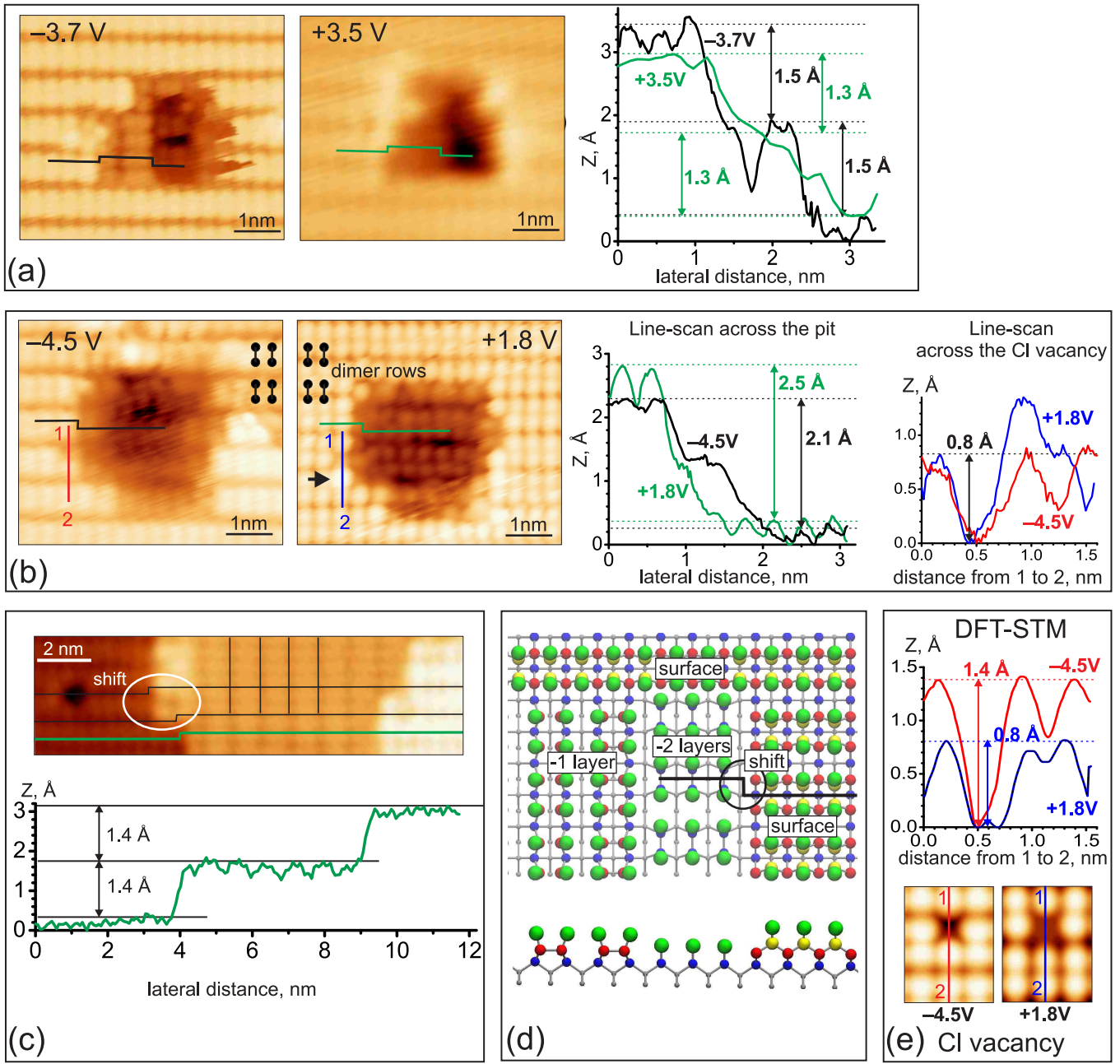}
    \caption{(a) STM images of filled (U$_s =-3.7$\,V, I$_t$ = 1.0\,nA) and empty (U$_s =+3.5$\,V, I$_t$ = 1.0\,nA) state of the pit created on Si(100)-2$\times$1-Cl and line-scan across the pit (drawn over top of chlorine atoms on the surface and bottom of the pit). (b) STM images of filled (U$_s =-4.5$\,V, I$_t$ = 1.0\,nA) and empty (U$_s =+1.8$\,V, I$_t$ = 1.0\,nA) state of another pit (the arrow indicates a "ball-like" defect) and line-scan across the pit and across the Cl vacancy at the edge of the pit. (c) Empty state STM image (U$_s =+2.5$\,V, I$_t$ = 1.0\,nA) of the surface area with two steps and line-scan across the steps. (d) Top and side view of the model showing the atomic structure of the Si(100)-2$\times$1-Cl surface with one and two silicon layers removed. Chlorine atoms are indicated by green circles, Si atoms belonging to the first, second, and third layers  are indicated by yellow, red, and blue circles, respectively. (e) Simulated STM images of the Cl vacancy and cross-section lines for filled ($-4.5$\,V) and empty ($+1.8$\,V) states.
 }
    \label{fig2}
\end{center}
\end{figure*}

Figures~\ref{fig2}a,b show atomically resolved STM images of two pits. According to Fig.~\ref{fig2}a, dimer rows at the first step of the pit are aligned perpendicularly to the direction of dimer rows on the terrace. Dimer rows at the bottom of the pit are parallel to the direction of dimer rows on the terrace and shifted by a half of the Si(100) lattice constant. The depth of the pit is equal to 2.6\,{\AA} and 3.0\,{\AA} for empty and filled states, respectively. The height of single step of the pit is equal to 1.3\,{\AA} and 1.5\,{\AA} for empty and filled states, respectively. Dimer rows of another pit (Fig.~\ref{fig2}b, empty states), are also parallel to the direction of dimer rows on the terrace and shifted by a half of the Si(100) lattice constant. The depth of this pit measured for empty  and filled  states is equal to 2.5\,{\AA} and 2.1\,{\AA}, respectively (Fig.~\ref{fig2}b).

Figure~\ref{fig2}c presents the STM image of the Si(100)-2$\times$1-Cl surface with two atomic steps and the line-scan across the steps. The step height is equal to 1.4\,{\AA} (Fig.~\ref{fig2}c), being close to the step height of the pit in Fig.~\ref{fig2}a (1.3--1.5\,{\AA}). Note, the measured depth of the pit is higher if the bottom of the pit is imaged with atomic resolution (Fig.~\ref{fig2}a,b). Thus, loss of atomic resolution resulted in the decrease of the depth from $\approx$3.0\,{\AA} to $\approx$2.6\,{\AA} and from $\approx$2.5\,{\AA} to $\approx$2.1\,{\AA} for the first (Fig.~\ref{fig2}a) and the second (Fig.~\ref{fig2}b) pit, respectively. Therefore, if the tip is not sharp enough, the depth of the pit is underestimated. Thus, we believe that the depth of 1.5\,{\AA} and 3.0\,{\AA} of the pit corresponds to the height of a single (1.4\,{\AA}) and double (2.8\,{\AA}) steps, respectively.

Analysis of dimer rows directions in the pits can give additional information. Indeed, the removal of one layer of silicon reveals the dimer rows with the perpendicular orientation (Fig.~\ref{fig2}c). This case is obviously realized on each atomic step where the change of the direction of the dimer rows occurs. The removal of two silicon layers is equivalent to a double step on the Si(100) surface, and according to the STM image in Fig.~\ref{fig2}c is accompanied by the shift of dimer rows by a half of the Si(100) lattice constant. Figure~\ref{fig2}d shows model drawings explaining the change of dimer rows direction on the steps. Thus, the structures of dimer rows observed in the pits (Fig.~\ref{fig2}a,b) clearly show that the pits have the depth of two steps. It also means that the bottom of the pits in Fig.~\ref{fig2}a,b is covered with chlorine. Therefore, there are no doubts that our nano-patterning algorithm results in the desorption of silicon atoms.

Edges of the created pits contain some defects. At edge of the pit in Fig.~\ref{fig2}b, there is a "ball-like" defect that is slightly brighter than the background chlorinated Si dimers in empty state STM image. The "ball-like" defect is not a Cl vacancy or Si adatom, since this defect is not visible in the filled state STM image. We suppose this "ball-like" defect to be a charge state of the chlorine atom, which appears only at positive bias. There is a dark spot visible in both polarities nearby the "ball-like" defect. According to the line-scan across the edge of the pit (Fig.~\ref{fig2}b), the height difference of the dark spot is about 0.8\,{\AA} (for both bias polarities). Figure~\ref{fig2}e shows the simulated STM images of a single chlorine vacancy (in accordance with previous experiments \cite{1998Lyubinetsky} and calculations \cite{2004Lee}). The height difference in the simulated STM images of the vacancy is about 1.4\,{\AA} and 0.8\,{\AA} for filled and empty states, respectively (Fig.~\ref{fig2}e). The height difference of the dark spot in the experimental empty state STM image (Fig.~\ref{fig2}b) is consistent with corresponding height difference of chlorine vacancy (Fig.~\ref{fig2}e). Also, STM images of the dark spot (Fig.~\ref{fig2}b) and Cl vacancy look similar (Fig.~\ref{fig2}e) for both polarities. The lack of atomic resolution in the experimental filled state STM image (Fig.~\ref{fig2}b) could lead to an underestimation of the depth of the vacancy, which explains the deeper vacancy profile in the simulated filled state STM image (Fig.~\ref{fig2}e). Thus, data presented in Fig.~\ref{fig2}b,e indicate the absence of chlorine atoms on the edge of the pit. In principle, chlorine may also be missed inside the pits. A deficit of chlorine in the pits can be explained both by the increase of the total surface area (taking into account the walls of the pits) and chlorine removal (into vacuum or onto the tip or onto the surface) during the process of pit creation.

To summarize this section, we demonstrated that the STM pulse (linearly increased from $+3$ to $+4$\,V for 2\,s at a current of 1\,nA and at a certain distance from the tip to the surface) leads to the formation of etching pits on the Si(100)-2$\times$1-Cl surface with silicon atoms removed. Unfortunately, a direct comparison of our results with data obtained by Dwyer et al. \cite{2018Butera}, who observed a removal of chlorine atoms only, is difficult due to different parameters used in our works. In particular, for positive single pulses, the pulse  $+4.25$\,V at room temperature was shorter by about two orders of magnitude, and at 4\,K during the pulse ramped from $-3.5$\,V to $+10.0$\,V the current was at least ten times less than in our work. Accordingly, in both cases the STM tip influence on the Si(100)-2$\times$1-Cl surface was much less than in our case.

\subsection{Mechanism of pit creation}

Desorption of atom can be viewed as atom escaping from the potential well of the surface to the potential well of the tip or into the vacuum. To solve this problem, we need to estimate the activation barriers for removing different atoms (or silicon-adsorbate compounds) from the Si(100)-2$\times$1-Cl surface. According to our DFT calculations, the activation barrier for Cl atom removing is equal to 4.3\,eV. The energy for the Cl$_2$ and SiCl removal was taken as the difference between the energy of initial and final states and are equal to 5.0\,eV and 6.3\,eV, respectively. (Taking into account the  barriers between initial and final states can lead to an increase in the energy required for desorption. In our case, it is not important because the obtained values already exceed the voltage applied for the pit creation). Desorption of \ce{SiCl2} takes place in two stages: the \ce{SiCl2} formation and the \ce{SiCl2} desorption. The activation energy of the entire process is 4.6\,eV \cite{1997Wijs} (the activation energies of the first and second stages are 2.1\,eV and 3.2\,eV, respectively). Single bonds of silicon atoms can also be broken, because the Si--Si bond energy is equal to  2.4\,eV \cite{2010Yuan}. Note that the electric field is not taken into account in these calculations; however, it can influence the height and the width of the potential barrier between the tip and the sample \cite{2015Minato}.

In our experiments, we use the voltage pulse of 3--4\,V which is enough to break Si--Si and Si--Cl bonds, but is not enough for desorption of \ce{Cl2} or \ce{SiCl}. Also the energy transferred by two electrons from the tip may be enough for the \ce{SiCl2} desorption, but the intermediate state should have a sufficiently long lifetime. In pits deeper than one silicon layer, a number of Si atoms removed is greater than a number of Cl atoms; therefore, \ce{SiCl2} compounds cannot be the main desorption product. Thus, the successive removal of Cl and Si atoms from Si(100)-2$\times$1-Cl seems to be the most likely in our experimental conditions.

One of the possible mechanisms of pit creation on the Si(100)-2$\times$1-Cl surface is field-induced extraction of Cl and Si atoms. If positive voltage is applied to the sample, the desorption of positively charged Si ions should occur. Evaporation of silicon clusters of 20–50 atoms (mainly as Si$^{+}$ ions) was reported in field ion microscopy (FIM) experiments in vacuum at and below 78\,K \cite{1979Sakurai}. According to the calculations, the lowest field required for silicon evaporation is 2.76\,V/{\AA} for Si$^{+}$ ions and 3.28\,V/{\AA} for Si$^{2+}$ ions  (at 300\,K, 1\,s)  \cite{1992Miskovsky}. In FIM, Si$^{2+}$ ions were evaporated by the field of 3.8\,V/{\AA} \cite{2005Tsong}. Although these fields are too high to be realized in our STM experimental conditions, the extraction of atoms from the surface by STM tip requires lower fields than in FIM \cite{1991Tsong, 1991Lyo}. For example, the removal of silicon atoms from Si(100)  \cite{1993Kobayashi, 1994Kobayashi} and Si(111) \cite{1991Lyo, 1993Kobayashi_Science} surfaces without adsorbate at both positive and negative voltages ($|U|>$2\,V)  was attributed to field evaporation. On chlorinated Si(111) surface, a positive bias voltage pulse (4--6\,V) applied to the sample resulted in field desorption and readsorption of chlorine \cite{1994Baba64, 1994Baba12} and Si evaporation  \cite{1994Baba65}. Thus, we also suggest that the desorption of atoms from the Si(100)-2$\times$1-Cl surface by the STM pulse can be induced by the field.

However, during the voltage rump the current runs bigger than 10\,nA, which means that electron-stimulated desorption (ESD) cannot be excluded as well. (Unfortunately, our electronic system does not allow measuring currents above 10\,nA.) The mechanisms of ESD from the Si(100)-2$\times$1-Cl surface under the STM tip action have not been studied theoretically until now but there is a substantial number of works investigating hydrogen removal from Si(100)-2$\times$1-H by the STM tip \cite{1995Shen, 2010Yuan, 1997Kratzer, 1998Stokbro, 1998Foley, 1999Syrykh, 2003Soukiassian}. We will use the case of Si(100)-2$\times$1-H to shed more light at possible mechanisms of the electron-stimulated desorption from Si(100)-2$\times$1-Cl.

There are two different mechanisms of ESD on Si(100)-2$\times$1-H resulting in the break of the bond between the atom and the surface \cite{2009Walsh}. First, the transition of an electron from the tip to the unoccupied antibonding  state  ($\sigma^{\ast}$) of a surface atom, and second, the excitation of the bonding state ($\sigma$) with further transition of an electron to the antibonding state. The latter mechanism operates at 5--6\,eV, which corresponds to the minimum energy difference between the peaks of the occupied and unoccupied states  \cite{1995Shen}, while the former ESD mechanism is considered to be dominant at 2--5\,V.

\begin{figure}[h]
\begin{center}
    \includegraphics[width=\linewidth]{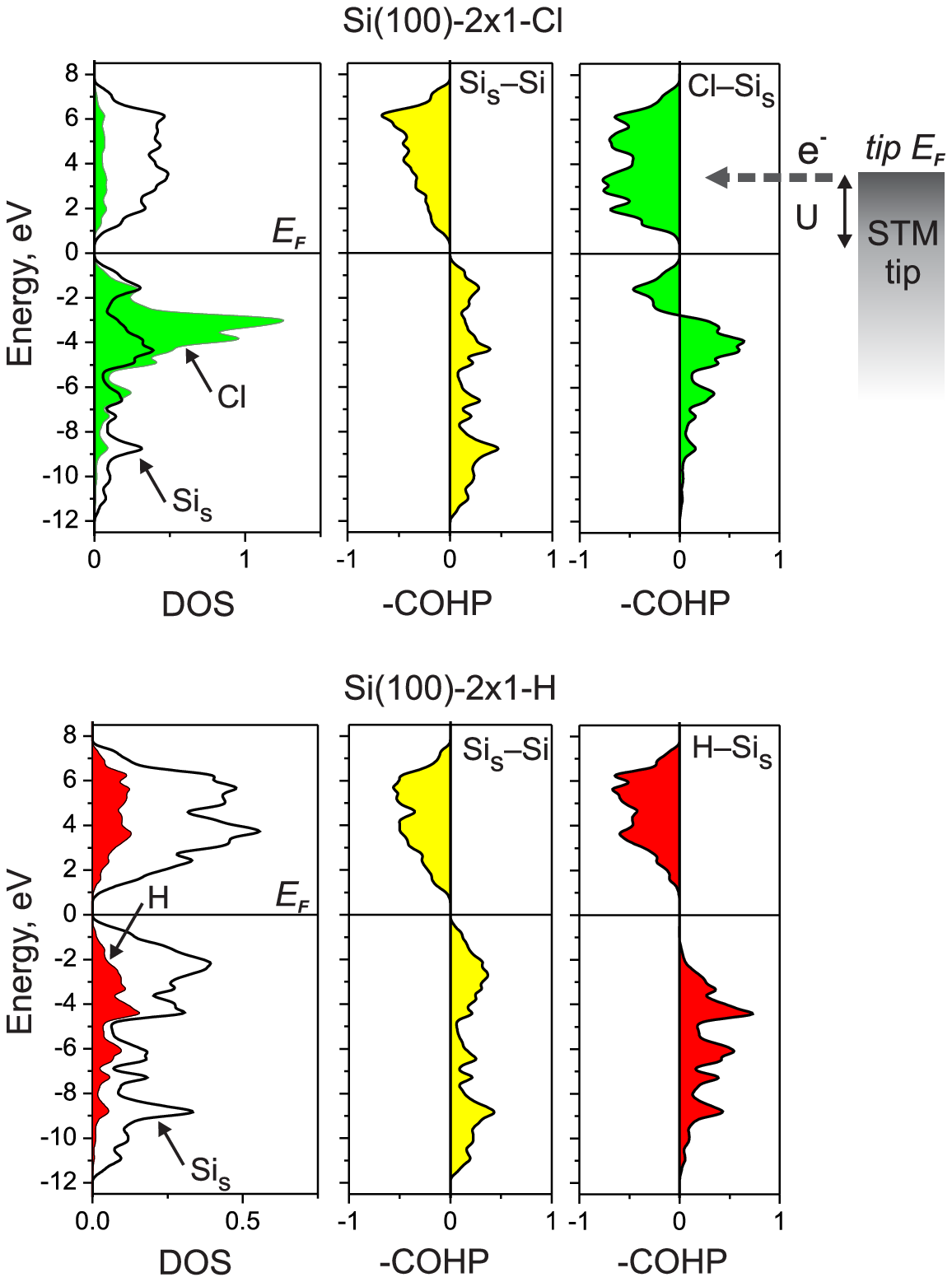}
    \caption{Density of states (DOS) and crystal orbital Hamilton population ($-$COHP) of the chlorine (hydrogen) and surface silicon atom Si$_s$ on Si(100)-2$\times$1-Cl (Si(100)-2$\times$1-H) ($E_F$ is the Fermi level). In the upper part, the  transition of electrons from the Fermi level of the tip to an antibonding unoccupied states is schematically represented.}
    \label{fig3}
\end{center}
\end{figure}

Figure~\ref{fig3} shows the density of states of a chlorine (hydrogen) atom and a surface Si$_s$ atom on a chlorinated (hydrogenated) Si(100)-2$\times$1 surface. All the unoccupied states (above the Fermi level) are antibonding  ($-$COHP$<0$ for $E>E_F$, see Fig.~\ref{fig3}), and the higher occupied states for Si(100)-2$\times$1-Cl are antibonding too.  Antibonding unoccupied states are spread from 1\,eV and higher with a first peak at about 2\,eV. Taking into account an upward band banding at a positive voltage pulse, it will require a little more than 2\,eV for an electron transition from the tip to the unoccupied antibonding state. Thus, the voltage pulse of 3 to 4\,eV may be sufficient for supporting the first ESD mechanism of chlorine and silicon desorption, and this energy fits in the voltage range used to create pits on the Si(100)-2$\times$1-Cl surface.

\begin{figure}[h]
\begin{center}
    \includegraphics[width=\linewidth]{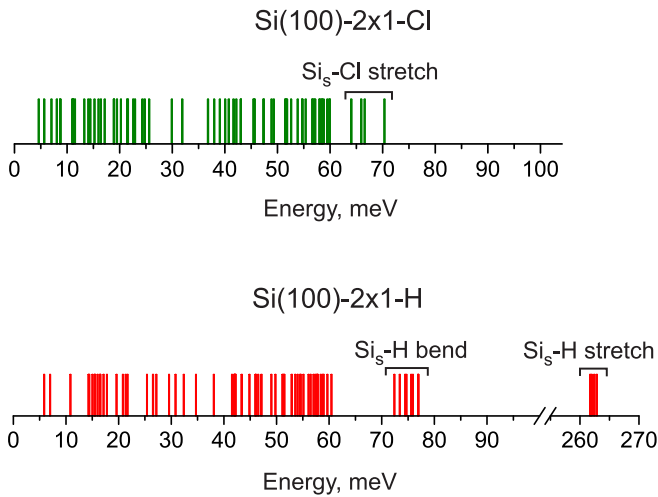}
    \caption{The calculated vibrational energy on the Si(100)-2$\times$1 surface with monolayer coverage of chlorine and hydrogen.}
    \label{fig4}
\end{center}
\end{figure}

Consider the second ESD mechanism --- excitation of the bonding state ($\sigma$) with further transition of an electron to the antibonding state. For hydrogen desorption from Si(100)-2$\times$1-H, electrons are inelastically scattered at the $\sigma^{\ast}$ antibonding state and excite vibrations of hydrogen on the surface \cite{1995Shen, 2017Moller, 2018Achal, 1998Stokbro, 2003Soukiassian}. The stretching mode of vibrations (Si$_s$--H) has energy about 260\,meV \cite{2006Andrianov}, which is far from other modes (Fig.~\ref{fig4}) and does not couple with them. The Si$_s$--H stretching mode has a very long lifetime (about $10^{-8}$\,s) \cite{1995Shen, 1995GuyotSionnest} and can therefore be excited by several electrons to the energy necessary for the desorption of a hydrogen atom (3.4\,eV). While incoherent multiple excitation (vibrational heating) requires up to fifteen electrons \cite{1995Shen,1998Stokbro}, an alternative mechanism of coherent multiple excitation \cite{2003Soukiassian} requires only two electrons. To determine the possibility of multiple vibrational excitation for Si(100)-2$\times$1-Cl, we calculated the vibrations of surface atoms and compared them with those of Si(100)-2$\times$1-H (Fig.~\ref{fig4}). The Si$_s$--Cl stretching vibrations have an energy of 66--68\,meV \cite{1994Gao}, which is close to the upper boundary of the phonon spectrum of surface silicon atoms (about 60\,meV). The Si$_s$--Cl bending mode lies within the silicon surface modes. Unlike hydrogen vibrations on Si(100)-2$\times$1-H, excited Si$_s$--Cl vibrations can relax due to the strong coupling with surface vibrations and should not have a long lifetime. Thus, in the case of Si(100)-2$\times$1-Cl, multiple excitation of the single vibration is unlikely. Nevertheless, single-electron excitation or coherent multiple excitation of bonds on the surface can have an impact on desorption at short vibrational lifetimes. Note, surface vibrations can induce a local heating effect on the surface under the tip to make it easier for the atoms to escape the potential well.

Summarizing this section, we believe that both field-induced and electron-stimulated desorption have an impact on the pit formation on the Si(100)-2$\times$1-Cl surface at a voltage of 3--4\,V. Electric field induced by the STM tip makes the tip-surface potential barrier narrower, and, therefore, plays an important role in Si and Cl atoms desorption. Taking into account that the electric field is not very high at our experimental conditions, we suggest that ESD is partially responsible for atoms escaping the potential well. Since the electric field and the current have a combined influence on desorption, it is not so easy to separate the different processes of ESD:  $\sigma$  or $\sigma^{\ast}$ bond excitation by elastic or inelastic electron scattering, thermal activation, or other processes (not discussed in the present paper). Nevertheless, we think that multiple incoherent excitation of the single vibration on a chlorinated surface is impossible due to short lifetimes of Si$_s$--Cl and Si$_s$--Si vibrations. It means that at 3--4\,V, desorption processes induced by the STM tip on chlorinated and hydrogenated Si(100) surfaces are different. We can also exclude desorption due to the van der Waals interactions between the atoms of the tip and the surface \cite{1997Bartels}, since the van der Waals interactions are too weak to lead to the desorption of several silicon layers. Mechanical effect of the tip on the sample is not considered here, since the tip does not move toward the surface during the pulse.

\section{Conclusions}

We have demonstrated that STM-based lithography with silicon layers removal can be realized on a chlorine-covered Si(100)-2$\times$1 surface at 77\,K. According to our data, a positive bias voltage pulse (3--4\,V) leads to the reproducible formation of pits by removing several layers of silicon.

We considered various mechanisms of desorption from the Si(100)-2$\times$1-Cl surface and can conclude that, while both electric field and electron stimulated desorption are involved in pit formation in various ways, field-induced desorption is more likely to dominate the process at our experimental conditions. We have confirmed that at 3--4\,V, the activated desorption of Cl and Si atoms is possible. Most likely, Cl and Si atoms desorb successive and not in the form of SiCl or \ce{SiCl2} compounds. We have found that multiple incoherent excitation of the single vibration is unlikely to cause Cl and Si atoms desorption on Si(100)-2$\times$1-Cl, in contrast to H desorption from a hydrogenated silicon surface.

We believe that new results presented in this paper open up new possibilities of the three-dimensional local etching with the STM lithography. For example, pits can be used as markers for STM patterning of nanoelectronic devices. The advantage of the proposed method of STM-lithography on the chlorine resist is the possible preservation of the crystalline structure of silicon in the created pits.

\section*{Acknowledgments}
The work was supported by Russian Science Foundation (grant 16-12-00050). We also thank the Joint Supercomputer Center of RAS for providing the computing power.

%\bibliography{T:/Manuscripts/Bibliography/Bibtex_Tania}
\bibliography{Manuscript_rev_arxiv}

\end{document}